\documentclass[aps,twocolumn,preprintnumbers,prx,amsmath, amssymb,amsfonts,superscriptaddress,floatfix]{revtex4}

\usepackage{amsfonts}
\usepackage{amssymb}
\usepackage{amsmath}
\usepackage{ulem}
\usepackage{color}
\usepackage{latexsym}
\usepackage{graphicx}
\usepackage{subfigure}
\usepackage{graphics}
\usepackage{floatflt}
\usepackage{epsfig}
\usepackage{overpic}
\usepackage{soul} 
\usepackage{varioref,xr-hyper} 
\usepackage[latin1]{inputenc}
\usepackage{mathbbol}
\usepackage{makecell}

\definecolor{blue}{rgb}{0.0,0.35,1.0}

\newcommand{\be}{\begin{equation}}
\newcommand{\ee}{\end{equation}}
\newcommand{\bea}{\begin{eqnarray}}
\newcommand{\eea}{\end{eqnarray}}

\def\lb{\label}

\def\ra{\rightarrow}

\def\bk{{\bf k}}

\newcount\bozza \bozza=0
\ifnum\bozza=1
\newdimen\shift \shift=-1.0truecm
\def\lb#1{%
{\label{#1}\rlap{\kern\shift{$\scriptstyle#1$}}}}
\else\def\lb#1{\label{#1}} \fi

\begin{document}

\title{Correlation-driven enhancement of pairing in a nematic Hund's metal}

\author{Angelo Valli}
\affiliation{Department of Theoretical Physics, Institute of Physics, Budapest University of Technology and Economics, M\H{u}egyetem rkp. 3, H-1111 Budapest, Hungary}
\affiliation{HUN-REN-BME-BCE Quantum Technology Research Group, M\H{u}egyetem rkp. 3., H-1111 Budapest, Hungary}
\author{Laura Fanfarillo}\thanks{Correspondence: laura.fanfarillo@cnr.it}
\affiliation{Istituto dei Sistemi Complessi (ISC-CNR), Via dei Taurini 19, I-00185 Rome, Italy}
\date{\today}

\begin{abstract}
Superconductivity and nematicity coexist in the phase diagram of many correlated systems, including iron-based superconductors. We investigate how Hund-driven correlations reshape boson-mediated superconductivity in a multiorbital nematic metal. We find that dynamical correlation effects beyond a quasiparticle-only description are essential to capture the robustness of superconductivity in the Hund regime. In the nematic phase, Hund correlations simultaneously enhance the orbital differentiation of the superconducting gaps and inhibit the most extreme nematic-driven orbital polarization and coherence collapse that would otherwise suppress pairing at strong coupling. A controlled cutoff analysis reveals a nontrivial, orbital-dependent buildup of the gaps, indicating that different frequency windows of the correlated spectrum contribute unevenly to pairing in the nematic Hund regime. This implies that pairing mechanisms with different characteristic boson energies can lead to distinct gap structures and trends.
\end{abstract}

\maketitle

\section{Introduction}
\label{sec:intro}
Electronic nematicity and superconductivity are deeply intertwined in correlated materials, in particular in iron-based superconductors (FeSC), where a tetragonal-to-orthorhombic electronic instability breaks the $x/y$ symmetry and superconductivity emerges out of this already anisotropic metallic state \cite{Fernandes_NatPhys2014, Fernandes_Nature2022}. An itinerant perspective captures much of the FeSC phase diagram in terms of low-energy electronic instabilities: superconductivity is described in terms of electrons coupled by the exchange of collective bosonic fluctuations (most notably spin fluctuations), while nematicity is as an electronic instability driven by spin- or orbital-fluctuation physics in the $B_{2g}$ channel \cite{Fernandes_Nature2022}. Since both instabilities are electronic in nature, their interplay can involve subtle competition/cooperation effects and strong sensitivity to low-energy structure.

Several analyses have been carried out in this direction within an itinerant framework. In~\cite{Chen_PRB2020}, using a minimal BCS description, it was shown that the relative orientation between the nematic Fermi-surface distortion and the anisotropy of the superconducting (SC) gap can affect whether nematicity competes or cooperates with superconductivity, with outcomes that can be unusually sensitive to how contributions beyond the immediate Fermi-surface shell are regularized. A different route was proposed in~\cite{Fernandes_PRL2013}, where nematicity couples strongly to superconductivity when $s$ and $d$ pairing channels are nearly degenerate, enabling symmetry-allowed $s$--$d$ mixing terms that can qualitatively reshape the phase diagram. Further studies focused on the vicinity of a nematic quantum critical point, where soft nematic modes can provide an extra retarded attraction and enhance pairing~\cite{Lederer_PRL2015, Klein_PRB2018}. While these approaches capture key aspects of the nematic--SC interplay, they are typically formulated within a low-energy quasiparticle (QP) description.

A distinctive feature of FeSC, however, is that the normal state out of which both nematicity and superconductivity develop is not a weakly correlated metal. By now it is well established that a wide range of FeSC are well described as Hund's metals~\cite{Fernandes_Nature2022, Georges_PhysToday2024, Capone_ContPhys2026}. In these systems, the Hund exchange, $J_H$, controls coherence scales, promotes orbital selectivity, and redistributes spectral weight over a broad energy range~\cite{Georges_AnnRevCM2013, DeMedici_Chapter2017, Capone_ContPhys2026}. In this perspective, revisiting the interplay between nematicity and boson-mediated superconductivity in a Hund-correlated metal becomes a natural and timely task.

The relevance of Hund-driven correlations for nematicity and superconductivity has been discussed in complementary ways as recently reviewed in \cite{Capone_ContPhys2026}. Slave-spin mean-field studies showed that, once the $xz/yz$ symmetry is explicitly broken, correlations strongly renormalize the nematic response and enhance orbital selectivity of the orbital QP weights $Z_\mu$~\cite{Fanfarillo_PRB2017,Yu_PRL2018}, while favoring nematic configurations that avoid large $xz/yz$ charge imbalance~\cite{Fanfarillo_PRB2017}. Going beyond a QP-only description, Dynamical Mean-Field Theory (DMFT) studies~\cite{Fanfarillo_PRB2023} showed that in a nematic Hund metal, the $xz/yz$ spectral differentiation becomes strongly frequency dependent and extends to high energies, in qualitative agreement with ARPES indications of nematic spectral-weight redistribution~\cite{Pfau_PRB2021_122, Pfau_PRB2021_FeSe}. Within DMFT this behavior was linked to a sign change of the real-part self-energy anisotropy at intermediate-to-high frequencies~\cite{Fanfarillo_PRB2023}. However, the connection between self-energy inversion scales and low-energy observables has not yet been explored systematically.
On the SC side, Ref.~\cite{Fanfarillo_PRL2020} addressed how boson-mediated pairing can coexist with strong Hund-driven correlations, following a physical picture inspired by the fulleride scenario in which an effective pairing channel remains operative in the presence of strong local Coulomb interactions~\cite{Capone_Science2002, Capone_RMP2009, Nomura_SciAdv2015}. Within this framework, it was shown that boson-mediated pairing in a tetragonal Hund metal is controlled by dynamical correlations, with Hund-induced spectral redistribution bringing partially incoherent weight into the pairing window and enhancing gap robustness. Related multiorbital studies also emphasize intertwined orbital selectivity and pairing tendencies upon varying filling and correlation regime~\cite{Marino_PRL2025}.
These results suggest that combining nematicity and pairing in a Hund metal poses a nontrivial challenge. In particular, nematicity can reshape which frequency sectors of the Hund-renormalized spectrum contribute effectively to the Cooper kernel. Early attempts to address nematic superconductivity within orbital-selective QP frameworks emphasized the role of orbital-dependent $Z_\mu$ factors in producing gap anisotropies~\cite{Kreisel_PRB2017, Hu_PRB2018}; an important open issue is whether such QP-based descriptions remain sufficient once the nematic Hund metal develops a strongly frequency-dependent orbital differentiation.

In this work, we explicitly address three guiding questions: (i) does Hund-assisted enhancement of superconductivity persist in the nematic state?
(ii) how does nematicity reshape the orbital differentiation of the gaps across Hund regimes?
(iii) to what extent does the pairing cutoff act as a spectral filter in a nematic Hund metal?
To answer them, we combine a DMFT description of a correlated nematic three-orbital model with a mean-field BCS treatment. We systematically vary the pairing cutoff $\omega_0$ to probe how restricting the frequency window entering the Cooper kernel reshapes the orbital gaps in the nematic Hund regime. We do not assume orbital-dependent pairing interactions: the coupling is taken to be identical in all orbitals, so the orbital and cutoff-dependent gap structure emerges entirely from the correlated nematic electronic propagators. Our goal is not a material-specific quantitative fit, but to map the generic phenomenology of superconductivity in a nematic Hund metal across a broad parameter range and identify robust correlation-driven trends.

We find that dynamical correlations beyond a QP approximation are essential for a robust SC solution in the Hund regime. This
confirms that, also in the nematic case, pairing is sensitive to the redistribution of low-energy incoherent spectral weight encoded in the
dynamical self-energies. Hund correlations play a dual role in the nematic superconductor: they amplify nematic orbital differentiation of pairing while preventing the extreme orbital-selective coherence collapse that would suppress superconductivity. Finally, the cutoff evolution of the orbital gaps is nontrivial and orbital-dependent in the nematic Hund regime, consistent with a pairing kernel sensitive to the frequency distribution of incoherent spectral weight rather than to coherent QPs alone. Because nematic differentiation in the Hund metal is itself strongly frequency dependent, different $\omega_0$ effectively select different nematic sectors of the spectrum, naturally accounting for the distinct gap hierarchies observed at different cut-offs.\\

\section{Model and method}
\label{sec:model}

In order to study the effect of electronic correlations at a reasonable computational cost, we consider a minimal model, already used in~\cite{Fanfarillo_PRL2020, Fanfarillo_PRB2023}, which accounts for the main features of the electronic structure of FeSC and for the electron--electron correlations effects. The total Hamiltonian reads
\begin{equation}
H = H_0 + H_{\text{nem}} + H_{\text{int}}.
\end{equation} 
$H_0$, the kinetic Hamiltonian, is given by a three-orbital tight-binding model adapted from Ref.~\cite{Daghofer_PRB2010}:
\begin{equation}
H_{0} = \sum_{{\bf k}\sigma} \sum_{\mu \nu }T^{\mu\nu}({\bf k})\, c^{\dagger}_{{\bf k}\mu\sigma} c^{\phantom{\dagger}}_{{\bf k}\nu\sigma} \ ,
\end{equation}
where $\mu,\nu$ are orbital indices for the $yz$, $xz$, and $xy$ orbitals, and $c^{\dagger}_{{\bf k}\mu\sigma}$ ($c^{\phantom{\dagger}}_{{\bf k}\mu\sigma}$) creates (annihilates) an electron in orbital $\mu$, with momentum ${\bf k}$ and spin $\sigma$.
The set of parameters chosen, reported in the Supporting Information, reproduces qualitatively the shape and orbital content of the Fermi surfaces typical of the FeSC family, namely two hole-like pockets composed mainly of $yz$-$xz$ orbitals at the $\Gamma$ point and two elliptical electron-like pockets formed by $xy$ and $yz/xz$ orbitals centered at the $X/Y$ points of the 1Fe Brillouin zone.

We account phenomenologically for nematic order by introducing an orbital splitting in the $xz/yz$ sector, 
\begin{equation}
H_{\text{nem}} = \eta \sum_{\bf k} \big(n_{yz}({\bf k}) - n_{xz}({\bf k}) \big) , 
\end{equation}
where $n_{\mu}({\bf k})$ is the number operator and $\eta>0$ the magnitude of the nematic perturbation which explicitly breaks the tetragonal $xz/yz$ degeneracy.
As shown in~\cite{Fanfarillo_PRB2017, Fanfarillo_PRB2023}, once local correlations are included, this minimal symmetry-breaking seed yields a strongly nontrivial nematic response, with momentum- and frequency-dependent modulations of the $xz/yz$ spectra.
We emphasize that we do not assume any specific microscopic mechanism for nematicity here, nor do we search for a spontaneous instability within the present minimal model; instead, we treat $\eta$ as an external control parameter and focus on how local correlations reshape the resulting nematic electronic structure and, in turn, the SC instability.

Local electronic interactions are included by considering the multiorbital Kanamori Hamiltonian, which parametrizes the electron--electron interactions in terms of a Hubbard-like repulsion $U$ and an exchange coupling $J_H$~\cite{Georges_Review2013}:
\begin{eqnarray}
H_{int}&=& \frac{U}{2}\sum_{i \mu \sigma} n_{i \mu \sigma} n_{i \mu \bar{\sigma}} 
 + \frac{U^\prime}{2}\sum_{\substack{i \mu \neq \nu \\ \sigma  \tilde{\sigma}}} n_{i \mu \sigma} n_{i \nu 
 \tilde{\sigma}} +\nonumber \\
 && + \frac{J}{2}\sum_{\substack{i \mu \neq \nu \\ \sigma  \tilde{\sigma}}} c^{\dagger}_{i \mu \sigma}  c^{\dagger}_{i \nu \tilde{\sigma}} c^{\phantom{\dagger}}_{i \mu \tilde{\sigma}} c^{\phantom{\dagger}}_{i \nu \sigma} +\nonumber \\
 && + \frac{J}{2}\sum_{i \mu \neq \nu \sigma} c^{\dagger}_{i \mu \sigma}  c^{\dagger}_{i \nu \bar{\sigma}} c^{\phantom{\dagger}}_{i \nu \bar{\sigma}} c^{\phantom{\dagger}}_{i \nu \sigma} 
\end{eqnarray}
where $n_{i \mu \sigma}=c^{\dagger}_{i \mu \sigma} c^{\phantom{\dagger}}_{i \mu \sigma}$ is the density operator. $U$ and $U'$ are the intraorbital and interorbital Hubbard interactions, $J_H$ is the Hund's coupling. We assume the system to be rotationally invariant, and thus $U'=U-2J_H$ \cite{Castellani_PRB1978}. 
We compute the nematic orbital spectral functions in the normal state by using the full orbital- and frequency-dependent DMFT self-energy $\Sigma_{\mu\mu}(i \omega_n)$, where $\omega_n$ is the $n$-th fermionic Matsubara frequency following the same procedure implemented in~\cite{Fanfarillo_PRB2023}. The orbital dependence of the self-energy leads to a self-consistent renormalization of the nematic splitting. The effect of correlations within a Fermi-liquid quasiparticle picture, encoded in the quasiparticle weight $Z_{\mu}$, can be extracted from the low-frequency behavior of the DMFT self-energy  $Z_{\mu}= (1 - \partial \Im \Sigma_{\mu\mu}/\partial \omega_n)^{-1}$. 
We use an exact diagonalization DMFT solver at zero temperature~\cite{Capone_PRB2007, Weber_PRB2012, Amaricci_CPC2022} at filling $n=4$ electrons in three orbitals per site, which reproduces the low-energy electronic structure and the correlated nematic normal-state phenomenology relevant to FeSC.

We study superconductivity {\it on top of} this correlated nematic background, i.e., Cooper pairing of fully dressed electronic propagators in the nematic phase. In doing so, we assume that an effective boson-mediated pairing channel remains operative in the Hund-correlated regime, rather than being completely suppressed by the local Coulomb repulsion. This assumption is consistent with the few available vertex-level studies of Hund systems, which indicate that dynamical correlations can reinforce spin-mediated pairing interactions despite the strong suppression of quasiparticle coherence (see Ref.~\cite{Capone_ContPhys2026} for a review).
We consider spin-singlet pairing in the orbital basis and neglect interorbital anomalous components, i.e., we assume an orbital-diagonal gap $\Delta_\mu \propto \langle d_{-\bk\mu\downarrow} d_{\bk\mu\uparrow}\rangle$. The pairing Hamiltonian is written as
\begin{equation}
H_{\rm SC}=-\sum_{\mu\nu} g_{\mu\nu}\,\Delta_\mu^\dagger \Delta_\nu .
\end{equation}
The BCS-like coupling $g_{\mu\nu}$ is taken local and momentum independent, with a constant amplitude within the energy window $|\omega|<\omega_0$ and zero outside it. We further assume the pairing to be diagonal in the orbital basis, $g_{\mu\nu}=g\,\delta_{\mu\nu}$, i.e., we neglect pair-hopping between different orbitals. In what follows we deliberately keep the pairing interaction equivalent in all orbitals even in the nematic regime, so that any $xz/yz$ nematic differentiation of the superconducting gaps would arise from the nematic electronic structure and its correlation-induced self-energies, rather than being imposed directly through orbital selective couplings. The self-consistent BCS equations can be written compactly in orbital space as
\begin{equation}
\sum_{\nu}\Big( (\hat g^{-1})_{\mu\nu}-\Pi_{\mu\nu}\Big)\Delta_\nu = 0 ,
\end{equation}
where $\Pi_{\mu\nu}$ is the particle-particle kernel evaluated using the correlated nematic Green's functions. Notice that although the interaction is intraorbital, i.e., $\hat g$ is diagonal, the gap equations are {\it not} decoupled in orbital space because interorbital hybridization in the one-particle Hamiltonian mixes orbital characters and generates off-diagonal contributions to $\Pi_{\mu\nu}$.
Unless stated otherwise, all SC results reported in this work are obtained by constructing $\Pi_{\mu\nu}$ from the DMFT Green's functions that include the full frequency-dependent self-energies $\Sigma_{\mu\mu}(i\omega_n)$. When we refer to the QP approximation, we instead build $\Pi_{\mu\nu}$ using a low-energy linearization of the self-energy, retaining only the orbital quasiparticle weights $Z_\mu$ and static energy shifts. In both schemes, the self-energy is taken from the correlated normal state and is not recomputed in the SC phase. Accordingly, $g$ should be understood as an effective pairing interaction, incorporating at a phenomenological level the renormalizations of the pairing vertex that are not computed explicitly here. The present construction is expected to be most appropriate when the SC energy scale is small compared with the electronic scales governing the correlated normal state, as is typically the case in FeSC.

Unless stated otherwise, we set $\omega_0=\infty$, i.e., we use an {\it uncut} frequency-independent pairing interaction. To mimic a retarded boson-mediated interaction, we then introduce a finite cutoff $\omega_0$, restricting the pairing kernel to $|\omega|<\omega_0$, which we use as a controlled knob to probe which portions of the correlated spectrum effectively contribute to pairing. We fix $g=2$~eV so that the resulting gaps are numerically well resolved across the parameter scan. Since our focus is on relative trends with $U$, $J_H$, $\eta$, and $\omega_0$, the absolute gap scale is not interpreted as material specific. 

The explicit derivation of the self-consistent gap equations and numerical implementation are reported in the Supporting Information.

\section{Results and Discussion}
\label{sec:results}

\subsection{Normal-state coherence in the tetragonal and nematic phase}
\label{sec:normal}

\begin{figure}[tbh]
\centering
\includegraphics[width=1.00\columnwidth]{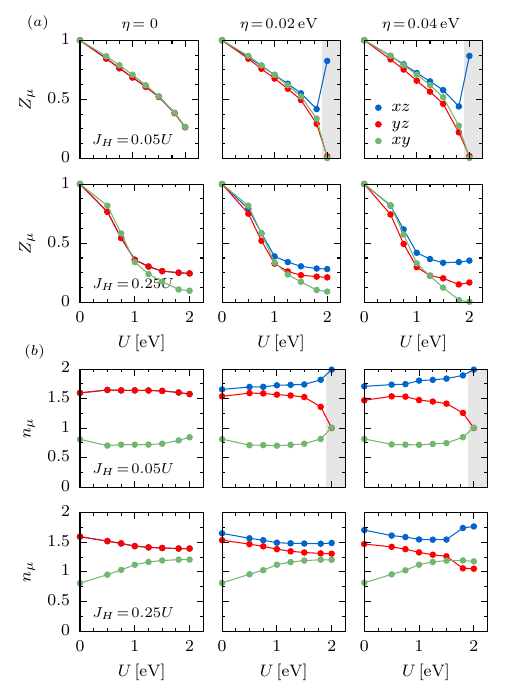}
\caption{{\bf Evolution of normal-state properties across the nematic-Hund crossover.} Normal-state orbital quasiparticle weights $Z_\mu$ (a) and occupations $n_\mu$ (b) as a function of $U$, for representative $\eta$ and $J_H/U$. (a) At low $J_H/U$ nematicity triggers a strongly orbital-selective coherence collapse, leading to an orbital-selective Mott (OSM) response (shaded region). At large $J_H/U$ the system remain metallic with finite $Z_\mu$ and a significant $xz/yz$ differentiation. (b) The corresponding occupations reveal that the OSM state arises from $\eta$-driven charge reorganization at low-$J_H/U$, a tendency substantially mitigated by large Hund's exchange.}
\label{fig:Z_overview}
\end{figure}

We start by characterizing the correlated nematic normal state from which superconductivity emerges.

Figure~\ref{fig:Z_overview}(a) summarizes the evolution of the quasiparticle weights $Z_\mu$ across the interaction range, for both $\eta=0$ and finite nematicity, $\eta>0$. 
In the absence of a nematic perturbation, $\eta=0$, we recover the typical phenomenology of the Hund's-metal crossover:
at low $J_H/U$ the QP weights $Z_\mu$ are suppressed monotonically as $U$ increases, consistent with a conventional correlated-metal evolution. For larger $J_H/U$ the suppression of coherence is more rapid at intermediate coupling and is followed by a long tail at larger $U$~\cite{DeMedici_PRL2011, Fanfarillo_PRB2015}, reflecting the emergence of Hund-metal physics. In this regime, the orbital decoupling promoted by $J_H$ leads to distinct orbital coherence scales~\cite{DeMedici_PRL2014}, and a clear differentiation between the $xz/yz$ and $xy$ channels sets in beyond the crossover.

Turning on nematicity, $\eta>0$ breaks the tetragonal $xz/yz$ degeneracy and reorganizes the coherence in an orbital-selective way. At low $J_H/U$, the nematic response becomes strongly amplified upon increasing $U$: the system undergoes an orbital-selective loss of coherence, leading to an orbital-selective Mott (OSM) regime, i.e., a selective collapse $Z_\mu\!\to\!0$ for two of the three orbitals. At large $J_H/U$, in contrast, $J_H$ strongly supports metallic behavior across orbitals: even in the presence of $\eta>0$ the orbital-selective coherence collapse is strongly softened, and metallicity is retained in all orbitals, with a sizable orbital differentiation of the $Z_{\mu}$.

The corresponding orbital occupations, shown in Fig.~\ref{fig:Z_overview}(b), rationalize this behavior. Already in the tetragonal phase, $\eta=0$, the occupations are not rigid: increasing $U$ induces a charge redistribution from the $xz/yz$ sector into the $xy$ orbital. This effect is weak at low $J_H/U$, but becomes pronounced in the Hund regime, where it drives the system toward a more uniform distribution across the three orbitals. 

For $\eta>0$, the nematic perturbation induces a charge transfer predominantly within the nematic pair $(xz,yz)$. At low $J_H/U$ this nematic-driven charge reorganization is strongly amplified as $U$ increases, driving two of the three orbitals toward (near) half filling while the other becomes correspondingly depleted or filled; this reorganization results in the OSM regime observed for the QP weights in Fig.~\ref{fig:Z_overview}(a). At large $J_H/U$, instead, the sharp nematic-driven polarization visible at low $J_H/U$ is substantially reduced: the $n_\mu$ remain at intermediate filling over the explored $U$ range, consistent with the tendency of Hund's exchange to disfavor strongly orbitally-polarized charge distributions, as also discussed in earlier slave-spin analysis~\cite{Fanfarillo_PRB2017}. This suppressed charge polarization provides the natural counterpart of the softened orbital-selective coherence collapse seen in Fig.~\ref{fig:Z_overview}(a).

Overall, the qualitative topology of the $Z_\mu$ versus $U$ curves is preserved under the nematic perturbation. The Hund coupling reshapes the degree of orbital selectivity by enhancing the orbital selectivity of the QP weights and suppressing the most extreme nematic-driven charge and coherence reorganization.\\

\begin{figure}[tbh]
\centering
\includegraphics[width=\columnwidth]{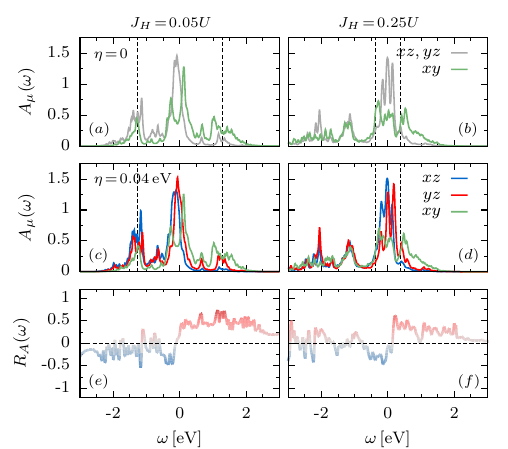}
\caption{{\bf Spectral evolution and orbital anisotropy in the nematic Hund's metal.} 
Orbital-resolved spectral functions $A_\mu (\omega)$ for $U=1.5$~eV at $J_H/U=0.05$ (a,c) and high $J_H/U=0.25$ (b,d). Vertical dashed lines indicate the effective Hubbard interaction $U_{\text{eff}}=U-3J_H$. 
(a,b) In the tetragonal case ($\eta=0$), increasing $J_H/U$ reduces the energy separation between coherent quasiparticle peaks and incoherent Hubbard bands. (c,d) In the nematic phase ($\eta=0.04$~eV), low $J_H/U$ results in a nearly rigid shift of the xz/yz spectra in opposite directions relative to the Fermi level ($E_F$). In contrast, the Hund regime (d) exhibits a strongly frequency-dependent $xz/yz$ imbalance. The $xy$ orbital is less sensitive to nematicity, while increasing $J_H/U$ suppresses its spectral weight near $E_F$. (e,f) The nematic anisotropy $R_A(\omega)$ shows a characteristic sign change at $E_F$ for low $J_H/U$, whereas the Hund regime develops complex finite-$\omega$ features, signaling a broad redistribution of the orbital imbalance.}
\label{fig:spectra_nem}
\end{figure}

To further characterize the correlated normal state beyond the QP weights, in Fig.~\ref{fig:spectra_nem} we discuss the orbital-resolved spectral functions $A_\mu(\omega)$ for representative interaction strength $U=1.5$~eV, comparing low and large $J_H/U$ focusing first on the tetragonal reference $\eta=0$ and then on the nematic case using $\eta=0.04$~eV. 

Already at $\eta=0$, the organization of spectral weight at finite energies is strongly affected by the Hund coupling. In a multiorbital Kanamori system, the characteristic energy scale for incoherent Hubbard-like features is not set by $U$ alone,
but by the lowest atomic charge-excitation energy $U_{\rm eff}$, which depends on filling~\cite{Georges_Review2013, Georges_PhysToday2024}. For the present filling $n\approx 4$ in the three-orbitals, i.e., away from half filling, the relevant excitation cost is $U_{\rm eff}\simeq U-3J_H$, so Hubbard-like features are expected to develop at energies of order $U-3J_H$. In the low-$J_H/U$ regime, this scale reduces essentially to $U_{\rm eff} \sim U$, and the orbital spectra exhibit a Mott-like organization, with Hubbard features developing away from a sharp low-energy coherent peak, as clearly visible in Fig.~\ref{fig:spectra_nem}(a). For $J_H=0.25\,U$, instead, the Hund coupling strongly reduces $U_{\rm eff}\simeq U-3J_H$; as a consequence the spectral functions do not exhibit a clear separation between coherent and incoherent features, and the redistribution of spectral weight mainly accumulates in a low-energy window around the Fermi level~\cite{Backes_PRB2015, Stadler_AP2018, Georges_PhysToday2024}, yielding the characteristic Hund-metal continuum and incoherent tail for all the orbitals as shown in panel (b). Comparing the strength of the renormalization among the three orbitals, the $xy$ channel shows the most evident modification at low-energy, with a strong broadening and a pronounced suppression of spectral weight close to the Fermi energy (pseudogap-like feature). This can be interpreted as a further consequence of the $xy$ filling being the closest to half-filling and correspondingly exhibiting not only the smallest quasiparticle weight $Z_{xy}$ among the three orbitals, but also the strongest low-energy spectrum suppression in the large-$J_H/U$ regime.

Turning on nematicity, $\eta$ differentiates the $xz$ and $yz$ spectra.
In the low-$J_H/U$ regime, panel~(c), the nematic perturbation predominantly produces a relatively rigid differentiation of the $xz/yz$ spectra around the Fermi level. In the Hund regime, in contrast, panel~(d), the nematic differentiation becomes strongly frequency dependent: the $xz$--$yz$ imbalance evolves non-trivially with $\omega$, with a frequency-modulated orbital redistribution of the nematic spectral weight. This is already visible in the $xz/yz$ line shapes and is emphasized by the spectral-asymmetry ratio:
\begin{equation}
R_A(\omega)=\frac{A_{yz}(\omega)-A_{xz}(\omega)}{A_{yz}(\omega)+A_{xz}(\omega)}\;,
\end{equation}
reported in panels~(e,f). At low $J_H/U$, $R_A(\omega)$ displays a clear sign change around the Fermi level, with a predominant $xz/yz$ character below/above $E_F$, quantitatively supporting the interpretation of the nematic spectra as a relatively rigid shift of $yz/xz$ spectral weight in opposite directions around $\omega=0$. In the Hund regime, in contrast, $R_A(\omega)$ develops additional positive ($yz$-dominated) regions at finite negative energies, where the low-$J_H/U$ case remained predominantly $xz$ dominated. This is consistent with the spectral line shapes in panel~(d), which show that part of the $yz$ spectral weight is redistributed to larger binding energies, producing an $yz$-dominated feature at negative $\omega$. Consistent with~\cite{Fanfarillo_PRB2023}, we verified that this frequency modulation is produced by a sign change of $\Re\Sigma_{yz}(i\omega_n)-\Re\Sigma_{xz}(i\omega_n)$ at intermediate frequencies (see Supporting Information), a self-energy fingerprint of the Hund-driven nematic spectral redistribution that underlies the cutoff sensitivity discussed below.
The $xy$ line shape appears comparatively less sensitive to $\eta$ than the $xz/yz$ sector at the level of this local spectral analysis. However, this does not automatically imply an absence of nematic effects in the $xy$ channel, since integrated observables (as probed by different experiments) could still display a measurable $\eta$ dependence.

\subsection{Superconductivity in the correlated nematic state}
\label{sec:sc}

\begin{figure}[tbh]
\centering
\includegraphics[width=1.00\columnwidth]{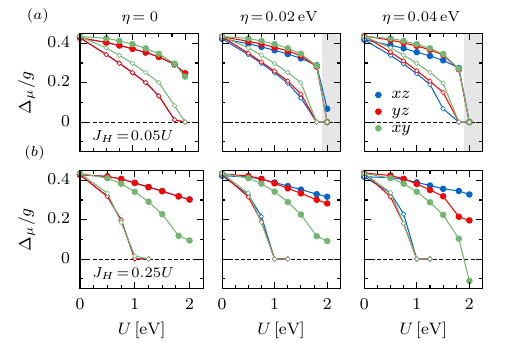}
\caption{{\bf Comparison between full dynamical and the quasiparticle approximation.} Orbital-resolved superconducting gaps $\Delta_\mu/g$ versus U (for $\omega_0=\infty$) for different nematicities $\eta$. Results from the full DMFT-based pairing kernel (filled symbols) are compared against the quasiparticle (QP) approximation (open symbols) for (a) low and (b) high Hund's coupling. The QP approximation systematically underestimates the gaps and artificially suppresses superconductivity at intermediate U, even while $Z_\mu$ is still finite. In contrast, the full DMFT solutions remain robust and sizable across the entire parameter range, despite the strongly suppressed $Z_\mu$ demonstrating that incoherent spectral weight is essential for pairing in the Hund regime.}
\label{fig:gap_DMFTvsQP}
\end{figure}

We now turn to superconductivity in the correlated nematic state. We start by explicitly benchmarking the role of dynamical correlations by comparing the solution of the BCS equations obtained using the full frequency-dependent DMFT self-energies with those obtained within a QP approximation. 
Figure~\ref{fig:gap_DMFTvsQP} shows that the QP treatment yields systematically smaller gaps than the full DMFT solution, and that this discrepancy becomes particularly pronounced at large $J_H$, where the QP solution collapses to $\Delta_\mu=0$ already at intermediate $U$, while the DMFT solution remains finite. 
Comparing Fig.~\ref{fig:gap_DMFTvsQP} with the quasiparticle weights in Fig.~\ref{fig:Z_overview} makes it clear that the robustness of the DMFT SC solution is not controlled by coherent quasiparticles $Z_\mu$. In the Hund regime, the gaps remain finite and sizable even when the corresponding quasiparticle weights are strongly suppressed. In particular 
at large $U$ and sizable $\eta$, the $xy$ orbital can become essentially incoherent ($Z_{xy}\!\to\!0$) while the SC solution still yields a finite $\Delta_{xy}$ across the whole explored nematic range.
This phenomenology persists across the full nematic range $\eta$ and directly extends to the nematic Hund metal the beyond-$Z$ robustness of the SC solution previously identified in the tetragonal case in~\cite{Fanfarillo_PRL2020}.

\begin{figure*}[tbh]
\centering
\includegraphics[width=1.00\textwidth]{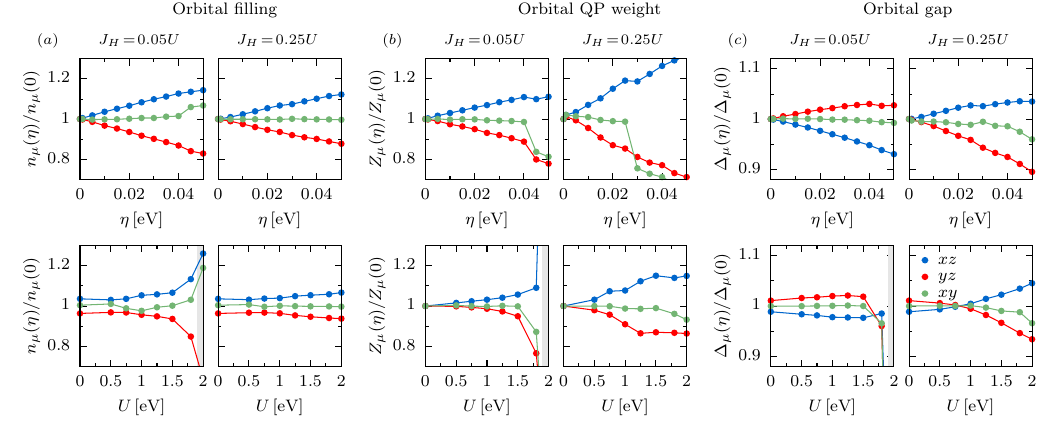}
\caption{{\bf Hund-driven stabilization of superconductivity and enhancement of orbital differentiation in the nematic phase.} Evolution of normal- and superconducting-state observables in the nematic phase, normalized to their tetragonal ($\eta=0$) values. Top row: $\eta$-sweep at fixed $U=1.5$~eV. Bottom row: $U$-sweep at fixed $\eta=0.02$~eV. (a) Normalized occupations $n_\mu(\eta)/n_\mu(0)$: at low $J_H/U$, nematic polarization of the $xz/yz$ orbitals is strongly amplified, whereas in the Hund regime, the occupations remain closer to their tetragonal values. (b) Normalized quasiparticle weights $Z_\mu(\eta)/Z_\mu(0)$: Hund's coupling enhances orbital differentiation while simultaneously preventing the transition into the OSM state. (c) Normalized gaps $\Delta_\mu(\eta)/\Delta_\mu(0)$: Hund correlations amplify the nematic differentiation of the superconducting response while inhibiting the OSM-driven collapse that quenches pairing at low $J_H/U$.}
\label{fig:nZDelta_eta0norm}
\end{figure*}

Beyond the overall enhancement, Fig.~\ref{fig:gap_DMFTvsQP} shows that in the Hund regime the SC response becomes more orbitally differentiated. To quantify this effect while factoring out the overall gap scale, we analyze $\eta$-normalized quantities
$\Delta_\mu(\eta)/\Delta_\mu(0)$ (and analogously $n_\mu(\eta)/n_\mu(0)$ and $Z_\mu(\eta)/Z_\mu(0)$), which isolate the nematic evolution within each orbital. Since at $\eta=0$ the $xz$ and $yz$ orbitals are symmetry equivalent, their deviation from unity provides a direct measure of the correlation-amplified nematic differentiation. 
Figure~\ref{fig:nZDelta_eta0norm} collects these three ratios to directly connect the nematic response of the SC gaps to the concomitant evolution of orbital fillings and coherence scales. Specifically, we show an $\eta$-scan at moderate coupling ($U=1$~eV) and a $U$-scan at fixed weak nematicity ($\eta=0.02$~eV). 

\paragraph*{Filling response and nematic-driven reorganization.}
Fig.~\ref{fig:nZDelta_eta0norm}(a) provides a compact view of how correlations reshape the nematic-induced redistribution of orbital fillings. In the $\eta$-sweep at $U=1$~eV (top row), the difference between small and large $J_H/U$ is relatively modest: increasing $\eta$ induces a moderate differentiation within the nematic pair, whereas the $xy$ filling remains weakly affected.
In the $U$-sweep at fixed $\eta=0.02$~eV (bottom row), instead, the two regimes separate clearly. At low $J_H/U$ the system develops a pronounced polarization of the nematic pair upon approaching the shaded OSM region, 
while at large $J_H/U$ Hund physics counteracts the nematic-driven orbital polarization and favors intermediate fillings, thereby suppressing the tendency toward a strongly orbitally unbalanced charge reorganization.

\paragraph*{Coherence response.}
Fig.~\ref{fig:nZDelta_eta0norm}(b) shows the corresponding coherence renormalizations. In the $\eta$-sweep at $U=1$~eV (top row), the Hund regime exhibits a pronounced nematic differentiation of coherence within the $xz/yz$ pair, by contrast, the low-$J_H/U$ case shows only a mild differentiation.
In the $U$-sweep at fixed $\eta=0.02$~eV (bottom row), the low $J_H/U$ regime is characterized by a sharp orbital-selective loss of coherence leading to an OSM state, whereas in the Hund regime this reorganization is substantially softened and the $xz/yz$ anisotropy evolves smoothly as $U$ increases, consistently with the corresponding filling trends.

\paragraph*{Gap response.}
Fig.~\ref{fig:nZDelta_eta0norm}(c) shows the normalized SC gaps. In the $\eta$-sweep at $U=1$~eV (top row) at low-$J_H/U$ the gaps remain very close to unity across the explored $\eta$ range, indicating a weak nematic selectivity of the SC response at moderate coupling. In the Hund regime, instead, the nematic selectivity is already clearly amplified at $U=1$ eV, whereas the $xy$ normalized gap stays comparatively close to unity in the same $\eta$ scan.
The bottom row (fixed $\eta=0.02$~eV, varying $U$) highlights most sharply the role of Hund-induced correlations. At low $J_H/U$, the three orbital gaps remain clustered near their $\eta=0$ reference values over most of the $U$-scan and then undergo a sharp suppression upon entering the OSM window. In this regime, the nematic-driven orbital-selective reorganization essentially drives a {\it collapse} of the SC solution at large $U$. In the Hund regime, by contrast, superconductivity remains robust up to the largest $U$ considered: the OSM-driven collapse is strongly softened by $J_H$, preventing the correlated nematic background from evolving into a configuration that would suppress pairing. As a result, finite gaps persist across the explored interaction range in {\it all} orbitals. At the same time, Hund correlations strongly {\it amplify} the orbital selectivity of the SC response already at moderate coupling: an evident $xz/yz$ differentiation of the gaps develops and continues to grow as $U$ is increased. The relative ordering of the orbital gaps correlates with the orbital differentiation of coherence. 

The analysis summarized in Fig.~\ref{fig:nZDelta_eta0norm} shows that Hund correlations play a dual role in the nematic superconductor: they amplify the nematic differentiation of coherence and pairing already at moderate coupling, while simultaneously suppressing the most extreme nematic-driven charge polarization and the associated OSM-like coherence collapse that would otherwise drive a collapse of superconductivity at large $U$.

\subsection{Cutoff dependence: spectral filtering in the nematic Hund metal}
\label{subsec:cutoff}
The nematic Hund's metal is characterized by a strongly frequency-dependent orbital differentiation of spectral weight, Fig.~\ref{fig:spectra_nem}(d). Because the $xz/yz$ imbalance changes both magnitude and sign across frequencies, restricting the pairing kernel to $|\omega|<\omega_0$ does more than simply truncate the available density of states: it selects different frequency sectors of the nematic imbalance itself. In this sense, the cutoff acts as a genuinely orbital-selective spectral filter, admitting distinct combinations of coherent and incoherent spectral contributions depending on its value. As a result, $\omega_0$ can bias the relative orbital content of the pairing kernel, making the cutoff evolution of the gaps particularly sensitive to the dynamical structure of nematicity in the Hund regime. 

\begin{figure}[tbh]
\centering
\includegraphics[width=1.00\columnwidth]{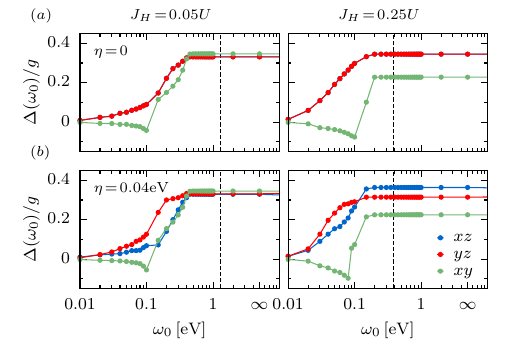}
\caption{{\bf Orbital-dependent gap buildup and hierarchy inversion.} Frequency cutoff dependence of the superconducting gaps $\Delta_\mu(\omega_0)/g$ at $U=1.5$~eV for $J_H/U=0.05,0.25$ in the tetragonal, $\eta=0$ (a) and nematic $\eta=0.04$~eV (b) cases. Dashed lines indicate the effective interaction scale $U_{\text{eff}}=U-3J_H$. In the Hund regime, dynamical correlations not only enhance the orbital differentiation of the asymptotic gap values ($\omega_0 \ra \infty$) but also induce a non-monotonic buildup. This results in a crossing of the xz and yz gap curves, showing that the gap hierarchy at low-energy cutoffs can be inverted compared to the full frequency-integrated results.}
\label{fig:gap_cutoff_rep}
\end{figure}
In Fig.~\ref{fig:gap_cutoff_rep} we report the cutoff evolution of the orbital gaps for representative parameters ($U=1.5$~eV, low-$J_H/U$ and Hund regime; $\eta=0$ and $\eta=0.04$~eV). Three robust qualitative messages can be extracted at the present resolution.
First, increasing $J_H/U$ enhances the orbital differentiation of the plateau values $\Delta_\mu(\omega_0\!\to\!\infty)$, consistent with the strong orbital selectivity of the correlated Hund regime discussed above. 
Second, the recovery toward the large-$\omega_0$ plateaus becomes faster upon increasing $J_H/U$. This trend is qualitatively consistent with the Hund-induced redistribution of incoherent spectral weight toward lower energies as $U_{\mathrm{eff}}\simeq U-3J_H$ is reduced. However, the cutoff scale governing the gap buildup is substantially smaller than $U_{\mathrm{eff}}$ and cannot be directly identified with it.
Third, the buildup with $\omega_0$ is nontrivial and orbital dependent: the orbital gaps grow in an orbital dependent way. In addition, the hierarchy of the gaps at small cutoffs need not coincide with the ordering of the plateau values. An orbital whose $\omega_0\!\to\!\infty$-gap is smaller, can exhibit a faster initial increase at small $\omega_0$, temporarily reducing or even inverting the gap hierarchy before the large-$\omega_0$ plateaus are reached. We further notice that the $xy$ component can change sign as $\omega_0$ is increased, indicating that the cutoff evolution is not simply a monotonic saturation but can involve cancellations in the orbital-resolved pairing kernel. 

\begin{figure}[b]
\centering
\includegraphics[width=1.00\columnwidth]{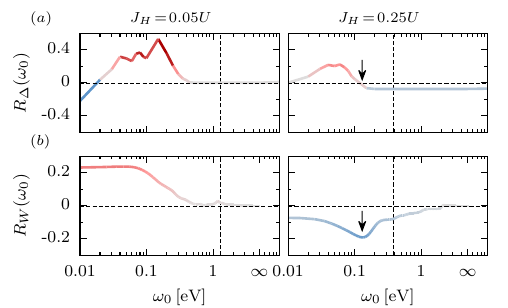}
\caption{{\bf Correlation between spectral weight redistribution and gap anisotropy.} Evolution of the gap and spectral anisotropies at $U=1.5$ eV and $\eta=0.04$ eV. Vertical dashed lines indicate the effective interaction $U_{\text{eff}}=U-3J_H$. (a) Gap anisotropy $R_\Delta(\omega_0)$: Increasing $J_H/U$ reduces the overall magnitude of $|R_\Delta(\omega_0)|$ and shifts its maximum toward significantly lower cutoffs. The arrow highlights the sign change ($R_\Delta(\omega_0)=0$) occurring specifically in the Hund regime. (b)  Integrated spectral anisotropy $R_W(\omega_0)$: in the Hund regime, the low-frequency sign of $R_W(\omega_0)$ is reversed compared to the low-$J_H/U$ case, and the peak anisotropy is shifted toward higher energy scales. Notably, the maximum of the spectral weight anisotropy (arrow) occurs at a frequency scale nearly coincident with the sign change in the gap anisotropy, suggesting that that high-energy dynamical correlations are the primary driver behind the inversion of the gap hierarchy.}
\label{fig:anisotropy}
\end{figure}

To move beyond the qualitative analysis of $\Delta_{\mu}(\omega_0)$ and better isolate the Hund-driven fingerprint in the cutoff response, 
we analyze the nematic $(xz/yz)$ gap anisotropy defined as
\begin{equation}
R_\Delta(\omega_0)\equiv
\frac{\Delta_{yz}(\omega_0)-\Delta_{xz}(\omega_0)}{\Delta_{yz}(\omega_0)+\Delta_{xz}(\omega_0)}.
\end{equation}
This dimensionless quantity factors out the overall gap scale and allows us to track the cutoff values at which the relative nematic differentiation is maximal or changes sign. In Fig.~\ref{fig:anisotropy}(a) we show $R_\Delta(\omega_0)$ at $U=1.5$~eV and $\eta=0.04$~eV for low and large $J_H/U$. In the low-$J_H/U$ case, $R_\Delta(\omega_0)$ exhibits a pronounced cutoff dependence, reaching a sizable anisotropy at intermediate $\omega_0$ before relaxing toward a near-zero plateau at large cutoffs.
In the Hund regime, $|R_\Delta(\omega_0)|$ is smaller, and its maximum occurs at substantially smaller $\omega_0$, indicating that the $xz/yz$ gap
differentiation is set by a different balance of low- versus higher-energy contributions admitted into the pairing window. Notably, the cutoff evolution in this case further displays a sign change followed by a long-lived finite plateau of $R_\Delta(\omega_0)$ at larger $\omega_0$, with opposite sign compared to the intermediate-$\omega_0$ anisotropy.

The spectral weight involved in the pairing window is given by $W_{\mu}(\omega_0)=\int_{-\omega_0}^{\omega_0}\! d\omega\, A_\mu(\omega)$ for each orbital. Following the same definition used for the gap anisotropy, we define the spectral $xz/yz$ anisotropy from
\begin{equation}
R_W(\omega_0)=\frac{W_{yz}(\omega_0)-W_{xz}(\omega_0)}{W_{yz}(\omega_0)+W_{xz}(\omega_0)}.
\end{equation}
Fig.~\ref{fig:anisotropy}(b) shows $R_W(\omega_0)$ for the same parameters. A clear qualitative difference emerges depending on the size of the Hund coupling. Compared to the low $J_H/U$, the Hund case displays an opposite low-$\omega_0$ sign of $R_W$ and a shift of the maximum $|R_W|$ toward larger cutoffs.
This demonstrates that Hund correlations reorganize the energy distribution of the nematic $xz/yz$ imbalance inside the pairing window. As a consequence, the orbital anisotropy of the spectra seen by the $|\omega|<\omega_0$ window becomes strongly $\omega_0$ dependent. In this perspective, the reduced, and strongly cutoff-dependent, gap anisotropy $R_\Delta(\omega_0)$ in the Hund regime is naturally consistent with the fact that the low-energy spectral-window anisotropy $R_W(\omega_0)$ has the opposite sign compared to the low-$J_H/U$ case. Notably, the cutoff at which $R_W(\omega_0)$ reaches its maximum magnitude occurs in nearly the same cutoff where $R_\Delta(\omega_0)$ changes sign, suggesting a direct link between the sign-changing spectral imbalance inside $|\omega|<\omega_0$ and the sign switch of the nematic gap anisotropy.

We attempted to further connect the cutoff evolution of $R_W(\omega_0)$ and $R_\Delta(\omega_0)$ to the crossing structure of $\Re\Sigma_{xz}-\Re\Sigma_{yz}$ in the Hund regime by tracking possible inversion scales as a function of $(U,\eta)$. With the present real-frequency resolution and the current sampling in $\omega_0$, we could not identify a robust monotonic trend that would allow us to extract a unique crossing-controlled cutoff scale across the explored parameter set. For completeness, we report the corresponding self-energy analysis in the Supporting Information.\\

It is worth noticing that all the strong orbital- and frequency-selective trends discussed in this Section, arise despite a completely orbital-independent pairing interaction, i.e. $g_{\mu\nu}=g\,\delta_{\mu\nu}$ with constant $g$. The orbital differentiation and cutoff dependence are exclusively generated by the correlated nematic one-particle sector (hybridization plus frequency-dependent self-energies) entering the Cooper kernel. 
In a realistic nematic system, however, the effective pairing interaction will generically break the $x/y$ symmetry as a consequence of both the anisotropy of the pairing fluctuations and correlation-induced vertex renormalizations. To test the robustness of our results against this additional source of orbital differentiation, we repeated representative calculations using $g_{yz}>g_{xz}$ and $g_{xy}=0$, a hierarchy compatible with an orbital-selective spin-fluctuation pairing scenario~\cite{Benfatto_npjQM2018, Fanfarillo_PRB2016, Chen_NatMater2019}. We find that the central conclusions of our analysis remain robust. In particular, large $J_H/U$ enhances the orbital differentiation of the superconducting gaps, prevents the OSM-driven suppression of superconductivity found at low $J_H/U$, and preserves a strongly orbital-dependent gap buildup as a function of the cutoff, with the $yz$ gap saturating at smaller $\omega_0$ than the $xz$ component. The detailed orbital-gap hierarchy and the occurrence of a crossing between the $xz$ and $yz$ gaps, however, are sensitive to the orbital structure of the pairing interaction and depend on the competition between the propagator-driven anisotropy identified here and the explicit anisotropy of the pairing vertex. When the latter is subdominant, the correlation-driven structure is retained, whereas strongly anisotropic $g_{xz/yz}$ can modify the gap hierarchy and remove the cutoff-induced crossing. The present calculation should be regarded as a robustness test rather than a quantitative estimate for a specific material. Establishing whether the same change occurs for realistic coupling strengths would require a microscopic determination of both the overall pairing scale and its orbital dependence.

\section{Conclusions}
\label{sec:conclusions}
We studied boson-mediated superconductivity developing on top of a correlated nematic Hund metal, extending to the nematic phase the DMFT+BCS framework introduced in Ref.~\cite{Fanfarillo_PRL2020}. As in that work, our analysis follows a physical picture inspired by the fulleride scenario, in which an effective boson-mediated pairing channel can remain operative in the presence of strong local Coulomb interactions~\cite{Capone_Science2002}. This assumption is supported by the few available vertex-level studies of Hund systems, which indicate that dynamical correlations can reinforce spin-mediated pairing interactions even when quasiparticle coherence is strongly suppressed (see Ref.~\cite{Capone_ContPhys2026} for a review). Our goal was not a material-specific fit, but to identify robust qualitative trends and disentangle quasiparticle-only effects from genuinely dynamical correlation physics. 
Three main conclusions emerge:

\paragraph*{(1) Superconductivity in the nematic Hund metal is supported by low-energy incoherent spectral weight.}
The quasiparticle treatment yields substantially smaller gaps, especially in the large-$J_H/U$ regime, and can suppress superconductivity
already at intermediate coupling, whereas the fully correlated DMFT-based solution remains finite and sizable even when quasiparticle coherence is
strongly degraded. This result confirms that, also in the nematic case, pairing is controlled by the dynamical redistribution of low-energy spectral weight encoded in the frequency-dependent self-energies as already pointed out by Ref.~\cite{Fanfarillo_PRL2020} in the tetragonal case.

\paragraph*{(2) Hund physics protects superconductivity in the nematic state.}
At low $J_H/U$, nematicity can strongly polarize the $xz/yz$ sector and trigger an orbital-selective loss of coherence that ultimately suppresses superconductivity at strong coupling. In the Hund regime, instead, Hund exchange suppresses this extreme nematic-driven charge polarization and coherence collapse, thereby stabilizing a correlated metallic state in which superconductivity remains robust. The resulting gaps become more orbitally differentiated and are enhanced relative to the low-$J_H/U$ case. Overall, Hund exchange makes superconductivity both more robust and orbitally selective in the nematic phase.

\paragraph*{(3) The pairing window matters: $\omega_0$ controls orbital gap buildup and anisotropy.}
Varying the pairing cutoff $\omega_0$ provides a controlled probe of which frequency sectors of the dressed spectrum effectively enter the Cooper
kernel. We find a nontrivial, orbital-dependent buildup of the gaps: the finite-$\omega_0$ evolution need not follow a simple monotonic approach to the $\omega_0\!\to\!\infty$ plateaus, and the hierarchy at small cutoff can differ from the asymptotic hierarchy. Diagnostics based on the nematic gap anisotropy $R_\Delta(\omega_0)$ and the spectral-window anisotropy $R_W(\omega_0)$ show that Hund correlations reorganize the energy distribution of the $xz/yz$ imbalance for $|\omega|<\omega_0$, yielding a reduced and strongly cutoff-dependent nematic gap anisotropy; notably, the maximum of $|R_W(\omega_0)|$ occurs near the cutoff where $R_\Delta(\omega_0)$ changes sign, pointing to high-energy dynamical correlations as a key ingredient behind the inversion of the gap hierarchy.

Our study shows that nematicity and Hund correlations combine to make the pairing problem intrinsically frequency selective and orbitally structured. Varying the pairing window reshapes not only the gap magnitude but also the orbital hierarchy and nematic anisotropy, implying that pairing mechanisms with different characteristic boson energies can realize distinct gap structures and nematic responses even when acting on the
same underlying Hund-correlated metal. This result provides a simple lens to interpret why different materials, or different tuning parameters within the same family, can exhibit distinct nematic gap structures despite sharing a similar Hund-correlated normal state.

\section*{Acknowledgment}
We thank M~Capone for valuable discussions. 
A.V. acknowledges support by the HUN-REN Hungarian Research Network through the Supported Research Groups Programme, HUN-REN-BME-BCE Quantum Technology Research Group (TKCS-2024/34).  

\bibliographystyle{apsrev}
\bibliography{NemSC}
\end{document}